\documentclass[aps,pre]{revtex4}
\usepackage[english]{babel}
\usepackage{amsmath}
\usepackage{epsfig}
\usepackage{dcolumn}
\newcommand{\tr}{\mathop{\rm tr}\nolimits}

\newcommand{\Pexp}{\mathop{\rm Pexp}\nolimits}

\newcommand{\inst}{\mathop{\rm inst}\nolimits}
\newcommand{\NP}{\mathop{\rm NP}\nolimits}

\newcommand{\eff}{\mathop{\rm eff}\nolimits}

\newcommand{\inputEps}[3]{

\centerline{\epsfxsize=#1 \epsfbox{#3} } \vskip 12pt
{\center\small{#2}} \vskip 18pt}
\begin{document}

\title{On the ground state energy of the inhomogeneous Bose gas}

\author{V.B. Bobrov and S.A. Trigger}

\address{Joint\, Institute\, for\, High\, Temperatures, Russian\, Academy\,
of\, Sciences, 13/19, Izhorskaia Str., Moscow\, 125412, Russia;\\
emails:\, vic5907@mail.ru,\;satron@mail.ru}

\begin{abstract}

Within the self-consistent Hartree-Fock approximation, an explicit expression for the ground state energy of inhomogeneous Bose gas is derived as a functional of the inhomogeneous density of the Bose-Einstein condensate. The results obtained are based on existence of the off-diagonal long-range order in the single-particle density matrix for systems with a Bose-Einstein condensate. This makes it possible to avoid the use of  anomalous averages. The explicit form of the kinetic energy, which differs from one in the Gross-Pitaevski approach, is found. This form is valid beyond the Hartree-Fock approximation and can be applied for arbitrary strong interparticle interaction. \\

PACS number(s): 05.30.Jp, 67.10.Fj, 67.85.Bc  \\

\end{abstract}

\maketitle

Experimental observation of the Bose-Einstein condensate (BEC) in ultracold gases of alkali metals [1] is a strong motivation for theoretical studies of weakly nonideal Bose systems. Due to the presence of magnetic moment the alkali metal atoms can be confined in magnetic traps. Ultralow temperatures requiring to form the BEC are achieved by use laser cooling which leads to evaporation of atoms with high energy from a magnetic trap (see [2] for more details). The ultracold gas obtained in such a way is rarefied and is characterized by strong inhomogeneity [3]. For these reasons, the Gross-Pitaevskii equation [4,5] corresponding to the "mean"\, field approximation and allowing the consideration of the effect of laser radiation [6] is widely used to describe the ultracold gas. Wherein, validity of the Gross-Pitaevskii equation is shown for the inhomogeneous system, containing a finite number of particles $N$ in an infinite volume $V$ [7]. The consideration of such system does not correspond to the thermodynamic limit transition ($N\rightarrow\infty$, $V \rightarrow\infty$, $N/V=const$). At the same time the derivation of the Gross-Pitaevskii equation for an inhomogeneous system in thermodynamic limit is based on the hypothesis of the "anomalous averages" existence (see [8]). As shown in [9-15] the
description of homogeneous systems with BEC using of anomalous averages is dubious.

An alternative approach proposed in the present paper is based on applying the conventional diagram technique of the perturbation theory to an equilibrium system in a large but finite volume [16] and on existence of the off-diagonal long-range order (ODLRO) in the one-particle density matrix. Such approach allows the self-consistent consideration of BEC by transferring to the thermodynamic limit [17] (see [18] for more details).
On this basis we establish in the present paper two new results: a) the explicit expression for kinetic part of the ground state energy of Bose system with BEC in an external field, which is valid for arbitrary strong interaction between particles, b) the ground state energy of the system under consideration in the Hartree-Fock approximation, which differs from one in the theory, based on anomalous averages.

We use ODLRO to describe an inhomogeneous system of bosons with zero spin and mass $m$, which is in a static external field defined by the scalar potential $\varphi^{(ext)}(\textbf{r})$. The Hamiltonian of such a system in the volume $V$ is written as
\begin{eqnarray}
\hat H=-\frac{\hbar^2}{2m}\int_V d^3 r \hat \psi^+(\textbf{r})\Delta_\textbf{r} \hat \psi (\textbf{r})+\frac{1}{2}\int_V d^3 r_1\int_V d^3 r_2 v(\textbf{r}_1-\textbf{r}_2) \hat\psi^+(\textbf{r}_1) \hat \psi^+ (\textbf{r}_2)\hat \psi(\textbf{r}_2)\hat \psi (\textbf{r}_1)+\int_V d^3 r \varphi^{(ext)}(\textbf{r})\hat \psi^+(\textbf{r}) \hat \psi (\textbf{r}),\label{A1}
\end{eqnarray}
where $\hat\psi^+ (\textbf{r})$  and $\hat \psi (\textbf{r})$   are, respectively, the field operators of creation and annihilation satisfying the bosonic commutation relations, and $v(r)$  is the pair interaction potential of particles. To determine the average energy $E_V=\langle\hat H\rangle_V$  of a rarefied weakly nonideal gas in the macroscopic volume  $V$, we can use the self-consistent Hartree-Fock approximation (see [19] for more details)
\begin{eqnarray}
E_V= \int_V d^3 r_1\int_V d^3 r_2 \left\{[-\frac{\hbar^2}{2m}\Delta_{\textbf{r}_2}\gamma_V (\textbf{r}_1,\textbf{r}_2)]\delta(\textbf{r}_1-\textbf{r}_2)+\frac{1}{2}v(\textbf{r}_1-\textbf{r}_2)n_V(\textbf{r}_1)n_V(\textbf{r}_2)+\right.\nonumber\\
 \left. \frac{1}{2}v(\textbf{r}_1-\textbf{r}_2)\gamma_V(\textbf{r}_1,\textbf{r}_2)\gamma_V(\textbf{r}_2,\textbf{r}_1)\right\}+\int_V d^3 r \varphi^{(ext)}(\textbf{r}) n_V (\textbf{r}).\label{A2}
\end{eqnarray}
Here $\gamma_V (\textbf{r}_1,\textbf{r}_2)]\equiv\langle\hat \psi^+(\textbf{r}_1) \hat \psi (\textbf{r}_2)\rangle_V$  is the sigle-particle density matrix, $n_V (\textbf{r})\equiv\langle\hat \psi^+(\textbf{r}) \hat \psi (\textbf{r})\rangle_V$  is the inhomogeneous density of the system under consideration in the macroscopic volume $V$, and angle brackets mean averaging with the grand canonical Gibbs ensemble.
As is known [20], the average values of physical quantities correspond to the state of thermodynamic equilibrium if the transition to the thermodynamic limit $V\rightarrow\infty$, $\langle\hat N \rangle_V \rightarrow\infty$, $\overline{n}= \lim_{V\rightarrow\infty}\langle\hat N \rangle_V/V=const$  is performed. Here $\langle\hat N \rangle_V=\int_V d^3 r n_V (\textbf{r})$  is the average total number of particles in the macroscopic volume $V$, and $\overline{n}$  is the average density of the number of particles in the thermodynamic limit [20].
 For transition to the thermodynamic limit in the calculation of the average energy $E=\lim_{V\rightarrow\infty} E_V $, we represent the functions  $\gamma_V (\textbf{r}_1,\textbf{r}_2)$  and $n_V(\textbf{r})$  in the form of the Fourier series
\begin{eqnarray}
\gamma_V (\textbf{r}_1,\textbf{r}_2)=\gamma_V ({\bf R}-\frac{\bf r}{2}, {\bf R}+\frac{\bf r}{2}) = \frac{1}{V} \sum_{\bf{p}} f_V(\textbf{p},\textbf{R})\exp (i \bf{p} \bf{r}),\label{A3}
\end{eqnarray}
\begin{eqnarray}
n_V (\textbf{r})=\gamma_V (\textbf{r},\textbf{r})= \frac{1}{V} \sum_{\bf{p}} f_V(\textbf{p},\textbf{r}),\label{A4}
\end{eqnarray}
where $f_V(\textbf{p},\textbf{R})$  is the inhomogeneous single-particle distribution function over momenta $ \hbar\textbf{p}$ , $\textbf{R}=(\textbf{r}_2+\textbf{r}_1)/2$, $\textbf{r}=\textbf{r}_2-\textbf{r}_1$. We note that, according to Eq. (3), the function $f_V(\textbf{p},\textbf{R})$  completely defines the average kinetic energy $\langle \hat K \rangle_V$ of inhomogeneous system in the macroscopic volume  $V$ [21]
\begin{eqnarray}
\langle \hat K_V \rangle=\int_V d^3 r_1\int_V d^3 r_2 [-\frac{\hbar^2}{2m}\Delta_{\textbf{r}_2}\gamma_V (\textbf{r}_1,\textbf{r}_2)]\delta(\textbf{r}_1-\textbf{r}_2)= \frac{\hbar^2}{2m V}\int d^3 R \sum_\textbf{p} \left(p^2-i \textbf{p} \nabla_\textbf{R}-\frac{1}{4}\triangle_\textbf{R} \right)f_V(\textbf{p},\textbf{R}). \label{A5}
\end{eqnarray}
We further take into account that in the region of ultralow temperatures in rarefied boson gas the BEC appears. The presence of the BEC manifests itself as the ODLRO for the function $\gamma_V (\textbf{r}_1,\textbf{r}_2)$  [22-24]. The existence of the ODLRO for the considering inhomogeneous system is defined as
\begin{eqnarray}
\lim_{|\textbf{r}|\rightarrow\infty}\gamma_V ({\bf R}-\frac{\bf r}{2}, {\bf R}+\frac{\bf r}{2}) = n^{BEC}(\textbf{R})\neq 0.\label{A6}
\end{eqnarray}
Here  $n^{BEC}(\textbf{R})$ is inhomogeneous local density of the number of particles in the BEC [25]. Therefore, according to (3), (6) the function $f_V(\textbf{p}=0,\textbf{R})= n^{BEC}(\textbf{R})V$  is a macroscopic quantity, which defines the existence of the BEC.

As a result, after transition to the thermodynamic limit in (3), (4), we find
\begin{eqnarray}
\gamma ({\bf R}-\frac{\bf r}{2}, {\bf R}+\frac{\bf r}{2})=\lim_{V\rightarrow\infty}\gamma_V ({\bf R}-\frac{\bf r}{2}, {\bf R}+\frac{\bf r}{2})=n^{BEC}(\textbf{R})+\gamma^{(over)}(\textbf{\textbf{r}},\textbf{R}), \label{A7}
\end{eqnarray}
\begin{eqnarray}
n(\textbf{r}) =n^{BEC}(\textbf{r})+n^{(over)} (\textbf{r}), \label{A8}
\end{eqnarray}
\begin{eqnarray}
\gamma^{(over)}(\textbf{\textbf{r}},\textbf{R})=\lim_{V\rightarrow\infty}\frac{1}{V}\sum_{\textbf{p}\neq 0} f_V(\textbf{p},\textbf{R})\exp (i \textbf{p} \textbf{r})=\int d^3 p f^{(over)}(\textbf{p},\textbf{R})\exp (i \textbf{p} \textbf{r}), \label{A9}
\end{eqnarray}
\begin{eqnarray}
n^{(over)} (\textbf{r})=\lim_{V\rightarrow\infty}\frac{1}{V}\sum_{\textbf{p}\neq 0} f_V(\textbf{p},\textbf{r})=\int d^3 p f^{(over)}(\textbf{p},\textbf{r}).\label{A10}
\end{eqnarray}
Here $\gamma^{(over)}(\textbf{\textbf{r}},\textbf{R})$  and $n^{(over)} (\textbf{r})$  are, respectively, the single-particle density matrix and the inhomogeneous density for "overcondensate"\, particles. The function $f^{(over)}(\textbf{p},\textbf{r})$ is the distribution over momenta $\hbar \textbf{p}$ under the condition $\textbf{p}\neq 0$.

According to Eq. (8), the average density of the number of particles is $\overline{n}=\overline{n}^{BEC}+\overline{n}^{(over)}$. For ultracold inhomogeneous gases which can be considered as weakly nonideal ones, almost all gas particles at temperature tends to zero ($T\rightarrow 0$)  are in the BEC: $\overline{n}^{BEC}\simeq\overline{n}$  [7,26]. Such an approximation was used, in particular, in deriving the Gross-Pitaevskii equation [4,5]. In the problem under consideration, this means that
\begin{eqnarray}
\lim_{T\rightarrow 0}\gamma(\textbf{\textbf{r}},\textbf{R})=n^{BEC}(\textbf{r});\qquad \lim_{T\rightarrow 0} n(\textbf{r})=n^{BEC}(\textbf{r}). \label{A11}
\end{eqnarray}
Using Eqs. (5), (11) and transferring to the thermodynamic limit in (2), we find the ground state energy as functional of the inhomogeneous BEC density
\begin{eqnarray}
E_0 [n^{BEC}]=\lim_{T\rightarrow 0} \lim_{V\rightarrow \infty}E_V=-\frac{\hbar^2}{2m} \int d^3 R \triangle_\textbf{R}n^{BEC}(\textbf{R})+ \int d^3 r \varphi^{(ext)}(\textbf{r})n^{BEC}(\textbf{r}) +\nonumber\\ \frac {1}{2}\int d^3 R \int d^3 r v(r) \left\{n^{BEC} ({\bf R}+\frac{\bf r}{2})n^{BEC} ({\bf R}-\frac{\bf r}{2})+n^{BEC} ({\bf R})n^{BEC} ({\bf R})\right\}. \label{A12}
\end{eqnarray}
The result obtained is based only on use of the ODLRO concept (4) for the single-particle density matrix within the self-consistent Hartree-Fock approximation. The hypotheses of the existence of anomalous averages in the considering method is not necessary.

Let us pay attention that, according to Eqs. (1), (5), (11), the first two terms on the right-hand side of relation (12), corresponding to the average kinetic energy and the energy in the external field potential, respectively, are exact, i.e. the form of these terms is also valid beyond the self-consistent Hartree-Fock approximation.

Since the inhomogeneous density $n^{BEC}(\textbf{r})$ is the nonnegative value we can represent it as $n^{BEC}(\textbf{r})=|\Phi(\textbf{r})|^2$, where $\Phi(\textbf{r})$ is so-called the wave function of BEC [8]. It is easily seen, that relation (12) for the energy of ground state of weakly nonideal Bose gas does not correspond to one from the equation of Gross-Pitaevskii [4,5]. The difference is drastically essential for the first term in the right part (12), which is the explicit form for the kinetic energy of an nonhomogeneous BEC (see (5)). The cause of this difference is conditioned by the applicability of the Gross-Pitaevskii equation only to the case of the finite number of bosons in infinite volume [7]. This system does not correspond to the description in thermodynamic limit which is considered in this paper.

According to the Gauss' theorem,
\begin{eqnarray}
\int d^3 R \triangle_\textbf{R}n^{BEC}(\textbf{R})=\oint d \textbf{S} \nabla_\textbf{R}n^{BEC}(\textbf{R}). \label{A13}
\end{eqnarray}
Therefore, if we neglect the surface integral over the infinitely-distant surface on the right-hand side of relation (13) the kinetic part of the ground state energy (the first term in (12)) for degenerate Bose gas equals zero: $\lim_{T\rightarrow 0}\lim_{V\rightarrow\infty}\langle \hat K \rangle_V=0$. This result corresponds to thermodynamic limit for arbitrary interparticle interaction potential. The physical sense of this result is clear: in inhomogeneous system of bosons BEC particles cannot contribute to the kinetic energy, as well as in homogeneous one. Formal application of the Gross-Pitaevskii equation for determination of inhomogeneous thermodynamics [8] leads to non-zero contribution to the kinetic energy and is invalid.

According (13), the ground state energy (12) for the inhomogeneous rarefied Bose gas in the self-consistent Hartree-Fock approximation is given by the relation
\begin{eqnarray}
E_0 [n^{BEC}]=\int d^3 r \varphi^{(ext)}(\textbf{r})n^{BEC}(\textbf{r}) +\frac {1}{2}\int d^3 R \int d^3 r v(r) \left\{n^{BEC} ({\bf R}+\frac{\bf r}{2})n^{BEC} ({\bf R}-\frac{\bf r}{2})+n^{BEC} ({\bf R})n^{BEC} ({\bf R})\right\}.  \label{A14}
\end{eqnarray}
Then, without loss of generality, we can apply the density functional theory [27] widely used in describing inhomogeneous electron systems (see [28] for more details) to the system under consideration. According to Eq. (14), the universal functional of the  inhomogeneous BEC density $F[n^{BEC}]$  in the self-consistent Hartree-Fock approximation is written as
\begin{eqnarray}
 F [n^{BEC}]=\frac {1}{2}\int d^3 R \int d^3 r v(r) \left\{n^{BEC} ({\bf R}+\frac{\bf r}{2})n^{BEC} ({\bf R}-\frac{\bf r}{2})+n^{BEC} ({\bf R})n^{BEC} ({\bf R})\right\}. \label{A15}
\end{eqnarray}
Then, within the Lagrange multiplier method taking into account the normalization condition $\int d^3 r (\textbf{r})n^{BEC}(\textbf{r})=\langle \hat N \rangle$, we find the equation for determining the function $n^{BEC}(\textbf{r})$
\begin{eqnarray}
\delta F [n^{BEC}]/\delta n^{BEC}(\textbf{r})+\varphi^{(ext)}(\textbf{r})=\mu, \label{A16}
\end{eqnarray}
where $\mu$  is the chemical potential of the system under consideration. Substituting Eq. (15) into (16), we find the integral equation
\begin{eqnarray}
\int d^3 r_1 v(|\textbf{r}-\textbf{r}_1|)n^{BEC}(\textbf{r}_1) + v_0 n^{BEC}(\textbf{r}) +\varphi^{(ext)}(\textbf{r})=\mu, \label{A17}
\end{eqnarray}
where $v_0=\int d^3 r v(\textbf{r})$.
The solution of Eq. (17) has a form $n^{BEC}(\textbf{r})=n^{BEC}(\textbf{r}, \mu, [\varphi^{(ext)}(\textbf{r}])$ in agreement with the Grand canonical ensemble requirements.

We note, that relation (17) as applied to the homogeneous degenerate Bose gas leads to the result $\mu=2  \overline{n} v_0 $ [11-13], in contrast with theory based on anomalous averages (see e.g. [8,29]).

The integral equation (17) can be simplified for slowly variable (in comparison with the characteristic length of the interaction potential) external field $\varphi^{(ext)}(\textbf{r})$. In this case
\begin{eqnarray}
\int d^3 r_1 v(|\textbf{r}-\textbf{r}_1|)n^{BEC}(\textbf{r}_1)=\int d^3 R v(|\textbf{R}|)n^{BEC}(\textbf{r}+\textbf{R})\simeq v_0 n^{BEC}(\textbf{r})+v_2 \Delta_\textbf{r}n^{BEC}(\textbf{r}), \label{A18}
\end{eqnarray}
where
\begin{eqnarray}
v_2=\frac{1}{3}\int d R R^2 v(R). \label{A19}
\end{eqnarray}
Therefore, in the weakly inhomogeneous case the approximate solution of Eq. (17) can be found from the differential equation
\begin{eqnarray}
v_2 \Delta_{\textbf{r}}n^{BEC}(\textbf{r}) + 2 v_0 n^{BEC}(\textbf{r})+\varphi^{(ext)}(\textbf{r})=\mu, \label{A20}
\end{eqnarray}	

Thus, the ground state energy of inhomogeneous BEC in thermodynamic limit is alternative to one following from Gross-Pitaevskii equation. This difference is conditioned by the application of anomalous averages for obtaining of GP equation. In the present consideration, based on ODLRO, the method of anomalous averages is avoided. The final choice between these alternative ways should be established on the basis of further theoretical investigations and comparison with the experimental measurements.

\section*{Acknowledgment}
The authors are grateful to A.G. Khrapak, A.A. Rukhadze, P.P.J.M. Schram and A.G. Zagorodny for helpful discussions.
This study was supported by the Russian Science Foundation, project no. 14-19-01492.\\

\end{document}